\documentclass[pra]{revtex4-2}

\usepackage{graphicx}
\usepackage{dcolumn}
\usepackage{bm}
\usepackage{xcolor}

\usepackage{CJKutf8}
\begin{document}
\begin{CJK*}{UTF8}{bsmi}

\title{Enhancement of broadband entangled two-photon absorption\\ 
by resonant spectral phase flips}
\author{Baihong Li (李百宏)}

\email{li-baihong@163.com}
\affiliation{%
Department of Physics, Shaanxi University of Science and Technology, Xi’an 710021, China
}%
\author{Holger F. Hofmann}%
 \email{hofmann@hiroshima-u.ac.jp}
\affiliation{%
Graduate School of Advanced Science and Engineering, Hiroshima University, Kagamiyama 1-3-1,
Higashi Hiroshima 739-8530, Japan
}%
\date{ received; revised; accepted; published}

\begin{abstract}
Broadband energy-time entanglement can be used to enhance the rate of two-photon absorption (TPA) by combining a precise two-photon resonance with a very short coincidence time. Because of this short coincidence time, broadband TPA is not sensitive to the spectrum of intermediate levels, making it the optimal choice when the intermediate transitions are entirely virtual. In the case of distinct intermediate resonances, it is possible to enhance TPA by introducing a phase dispersion that matches the intermediate resonances. Here, we consider the effects of a phase flip in the single photon spectrum, where the phases of all frequencies above a certain frequency are shifted by half a wavelength relative to the frequencies below this frequency. The frequency at which the phase is flipped can then be scanned to reveal the position of intermediate resonances. We find that a resonant phase flip maximizes the contributions of the asymmetric imaginary part of the dispersion that characterizes a typical resonance, resulting in a considerable enhancement of the TPA rate. Due to the bosonic symmetry of TPA, the enhancement is strongest when the resonance occurs when the frequency difference of the two photons is much higher than the linewidth of the resonance. Our results indicate that broadband entangled TPA with spectral phase flips may be suitable for phase-sensitive spectroscopy at the lower end of the spectrum where direct photon detection is difficult.
\end{abstract}

\maketitle

\section{Introduction}
\label{sec:intro}

It is well known that energy-time entanglement not only results in a linear dependence of the two-photon absorption (TPA) rate on photon flux \cite{Gea89,Jav90} but also in additional enhancements of the TPA rate \cite{You08,You09,You10,Oka10,Oka18a,Oka18b}. These quantum advantages have been demonstrated experimentally in atomic \cite{Geo95,Day04,Svo19} and molecular \cite{Lee06,Guz10,Guz13} systems, motivating a number of systematic studies of the influence of the specific level structure of these systems on the TPA rate \cite{Per98,Sal98,Leo13,Svo18,Svo19,Sch16a,Sch16b,Sch17a,Sch17b,Sch18,Shaul20,Lutz21,Raymer21,Ric10,Ros13}. However, there is still considerable debate concerning the reliability of theoretical predictions with regard to specific molecular systems \cite{Tab21,Par21,Raymer1,Raymer2,Raymer3,Raymer4}. The main conclusion that can be drawn from these studies seems to be that the actual enhancement depends on the specific level structure of the intermediate levels. The physical mechanism of entanglement-enhanced TPA is more complicated than the original idea suggested, and it is interesting to ask whether it is possible to introduce a new experimental method that would allow us to characterize the dependence of TPA enhancement on the spectrum of intermediate levels in a sufficiently simple and systematic manner. Specifically, we will focus on the possibility of modifying a broadband entangled state to maximize the off-resonant contributions to the TPA rate, resulting in an enhancement of TPA that makes optimal use of the broad range of frequency differences between the two photons in a broadband entangled state.

In the present paper, we point out that the theoretical description of TPA includes a very broad spectral feature that is anti-symmetric around the resonant frequency. The contribution of this anti-symmetric feature to the TPA rate can be enhanced significantly by introducing a phase flip in the spectrum of the broadband entangled two-photon state. Although the same phase flip reduces the symmetric contributions that are responsible for resonant absorptions, the enhancement of off-resonant absorption by a resonant phase flip tends to outweigh the loss of resonant TPA contributions. Our theoretical analysis is based on a general description of the TPA process where intermediate resonances are represented by Lorentz lines. We can then characterize the effects of a phase flip at a specific frequency for each of the intermediate resonances. The effect of the phase flip on a single intermediate resonance shows the expected resonant enhancement, which can be evaluated in terms of the ratio of the TPA rate relative to the TPA rate without the phase flip. We find that the resonant enhancement depends on the distance between the resonant frequency and the center of the spectrum, where both input photons have the same frequency. This result indicates that spectral phase flips could be used to identify intermediate resonances at the low end of the two-photon spectrum without the need for direct photon detection in that frequency region.

In systems with more complicated level structures, quantum interferences between the different intermediate levels may modify the effects of resonant enhancements. We have studied the effects of interferences between two levels with regard to both constructive and destructive interferences. When the linewidths are narrow, the resonant enhancements are modified only slightly. However, interference modifies the contributions to the TPA rate for entangled photons without a phase flip. In particular, two contributions can be completely cancelled by destructive interference. In this case, the phase flip is needed in order to activate TPA involving these intermediate levels. Spectral phase flips can thus be used to characterize interference effects between different resonances by contrasting the TPA observed when the phase flip is between the two resonances with the TPA observed without a phase flip.

The results of our analysis show that spectral phase flips can be used to characterize intermediate resonances in TPA using broadband entangled photons. Interestingly, the sensitivity of this method increases at the lower and higher ends of the spectrum, indicating that spectral phase flips may be a useful alternative to direct detection of radiation at these ends of the spectrum. It is also worth noting that the phase flip method can be applied to enhance TPA without any prior knowledge of the level structure of a material, simply by adapting the frequency of the phase flip to the experimentally observed resonances. Spectral phase flips may thus serve as a practical and uncomplicated method of adapting the phase dispersion of broadband entangled light to the spectral features of a specific material.

The rest of the paper is organized as follows. In section \ref{sec:phaseflip}, we give a brief introduction on the theoretical description of TPA in terms of the spectrum of intermediate levels and introduce a phase flip to the input state. In section \ref{sec:single}, we discuss the enhancement of the TPA rate achieved by the introduction of the phase flip for a single intermediate level. In section \ref{sec:Interference}, we discuss the interference effects between different resonances. Section \ref{sec:conclude} summarizes the results and concludes the paper.

\section{Effects of a spectral phase flip on the absorption of broadband entangled photons}
\label{sec:phaseflip}

The TPA process is described by the linear dynamics of a two-photon wavefunction and the level structure of the absorbing material. It is therefore possible to represent the probability of a TPA in terms of an inner product of the initial two-photon state and a maximally absorbed state representing the dynamics of the intermediate levels. If we consider only a single two-photon excited level, the absorption process will select photon pairs with a single sum frequency $\omega_+=\omega_1+\omega_2$. When coherent input light is used, only a small fraction of the incoming photon pairs will satisfy this condition. When entangled photons are used, the sum frequency of the two photons can be resonant with the two-photon transition between the ground state $|g>$ and the final state $|f>$, resulting in maximal absorption rates for the two-photon state. The rate of TPA then depends only on the overlap between the frequency difference wavefunction $\Phi(\omega_-)$ of the input photons with the frequency difference wavefunction of the optimally absorbed state $\Gamma(\omega_-)$ \cite{Li21,SPIE21},
\begin{equation}
\label{P_TPA}
P_{\mathrm{TPA}}= P_{\mathrm{gf}} \; \left|\int \Gamma(\omega_-) \Phi(\omega_-) d\omega_- \right|^2.
\end{equation}
The probability $P_{\mathrm{gf}}$ determines the transition probability when the overlap between the wavefunctions is one, representing the maximal achievable value of $P_{\mathrm{TPA}}$. The material response is encoded in the wavefunction of the optimally absorbed state. As shown in \cite{Li21}, it is possible to derive this wavefunction directly from the Hamiltonian of the electronic system, where different energy eigenstates appear as delta-like resonances,
\begin{widetext}
\begin{equation}
\label{eq:KKR}
\Gamma(\omega_-) = \sum_m 2 C_m \left(\pi \delta(\omega_-+\nu_m)+ \pi \delta(\omega_--\nu_m) + \frac{i}{\omega_-+\nu_m} - \frac{i}{\omega_--\nu_m} \right).
\end{equation}
\end{widetext}
In an idealized system, each intermediate level $m$ is an eigenstate of the Hamiltonian with an energy uncertainty of zero. The complex coefficients $C_m$ represent the transition matrix elements and the frequencies $\nu_m$ represent the difference between the single photon resonance of the level and the average frequency of the two absorbed photons, $\omega_{gf}/2$. Eq.(\ref{eq:KKR}) shows that every level $m$ contributes both a resonant part given by the delta functions at $\omega_-=\pm \nu_m$ and an off-resonant part related to the resonant contributions by Kramers-Kronig relations \cite{Li21}. $\Gamma(\omega_-)$ thus represents the Hamiltonian dynamics of the material response responsible for TPA.

In a realistic description of the material response, the Hamiltonian of the material system would have to include all interactions between the electronic system and other degrees of freedom. It should be noted that Eq.(\ref{eq:KKR}) can be applied to arbitrarily complicated Hamiltonians, including molecules with vibrational degrees of freedom resulting in the inclusion of the corresponding Franck-Condon factors. Unfortunately, it is difficult to handle the numerics of such a large number of intermediate states in an efficient manner. It may be interesting to consider possible simplified descriptions that can summarize clusters of intermediate states related to each other, but unfortunately, the discussion of a possible application to realistic molecules is beyond the scope of the present paper. Here, our main goal is the investigation of the physics of TPA with broadband entangled photons. We will therefore limit the following discussion to intermediate levels that couple to other degrees of freedom in a dissipative manner represented by a Lorentzian broadening of each level. The wavefunction describing the absorption process can then be expressed as
\begin{widetext}
\begin{eqnarray}
\label{Re-Im}
\Gamma(\omega_-)=\sum_m 2 C_m \left(\frac{1}{\gamma_m-i(\nu_m+\omega_-)} +  \frac{1}{\gamma_m-i(\nu_m-\omega_-)}\right),
\end{eqnarray}
\end{widetext}
where $\gamma_m$ is the linewidth of the resonance associated with the intermediate level $m$. The complex Lorentz lines in Eq.(\ref{Re-Im}) describe the characteristic phase dispersion of this resonance. The complex Lorentz lines in Eq.(\ref{Re-Im}) describe the characteristic phase dispersion of this resonance. It is possible to identify the real part of this response with the resonant contribution to the absorption and the imaginary part with the off-resonant contribution.
It may be worth noting that this separation of real parts and an imaginary parts corresponds to the separation shown in Eq.(\ref{eq:KKR}), which can be obtained from Eq.(\ref{Re-Im}) by taking the limit of $\gamma_m \to 0$ for all $m$. The resonant contribution is symmetric in the frequency difference $\omega_-$ around the resonance at $\pm\nu_m$ and the off-resonant contribution is anti-symmetric around $\pm\nu_m$. The spectrum of the off-resonant contribution is wider, confirming that it is the dominant contribution when the frequencies of the input photons do not match the resonances of the intermediate levels. However, the anti-symmetry of the off-resonant contribution means that they will tend to cancel out in the integral given in Eq.(\ref{P_TPA}) if the input state $\Phi(\omega_-)$ has the same phase for all frequency differences $\omega_-$ because the contributions have opposite sign for $\omega_-<\nu_m$ and for $\omega_->\nu_m$. In order to maximize the absorption associated with the imaginary part of $\Gamma(\omega_-)$, it is therefore useful to introduce a spectral phase flip in the spectral wavefunction $\Phi(\omega_-)$ of the input photons at $\omega_-=\nu_m$.

The introduction of a spectral phase flip at a specific frequency is a comparatively simple modification of a broadband entangled state that can be applied independent of the material properties of the absorber. Here, we consider the application of a phase shift of $\pi$ to all photons above a frequency of $\omega_{gf}/2+\delta_s$ emitted from a source of broadband entangled photons by an appropriate modulator. We would like to note that this is a particularly simple version of the more versatile pulse shapers that have been used to modify the wavefunction of entangled photons in previous experiments \cite{Day04,PRL2005,Dayan07,OL2013}. Using such techniques, it is possible to implement a broadband entangled state with a bandwidth of $b$ determined by the source of entangled photons and a phase flip at a variable frequency difference of $\omega_-=\delta_s$
\begin{equation}
\label{Phi_sym}
\Phi(\omega_-)= \left\{
\begin{array}{ccc} \displaystyle
-\frac{1}{\sqrt{b}} && \mbox{for} \hspace{0.2cm} |\omega_-|<\delta_s,
\\[0.5cm]
\displaystyle \frac{1}{\sqrt{b}} && \mbox{for} \hspace{0.2cm} \delta_s<|\omega_-|<b/2,
\\[0.5cm]
\displaystyle 0 && \mbox{for} \hspace{0.2cm} |\omega_-|>b/2,
\end{array} \right.
\end{equation}
where $0<\delta_s<b/2$. Note that the phase flip appears at both positive and negative values of $\omega_-$. This is a result of the bosonic symmetry of the photon wavefunction.  Also, we have chosen a negative sign for the wavefunction of the original input state, so that the wavefunction at $\delta_s=0$ is positive. Both $\delta_s=0$ and $|\delta_s|=b/2$ reproduce the original broadband entangled state, but the overall phase is opposite. The dependence of bandwidth and linewidth on the TPA rate without the phase flip has been studied in our previous work.


The phase flip in the dispersion of the state will change the contribution of each resonance to the integral in Eq.(\ref{P_TPA}) according to the difference between the resonance $\delta_s$ at which the phase flip is applied and the resonant frequencies $\nu_m$ and $-\nu_m$. It is possible to solve these integrals for each level $m$, where we distinguish the integral of the resonant contribution $A_m(\delta_s)$ from the integral of the off-resonant contribution $B_m(\delta_s)$. The rate of TPA is then expressed by the squared sum of these contributions,
\begin{eqnarray}
\label{P_AB}
P_{\mathrm{TPA}}(\delta_s)= \frac{P_{\mathrm{gf}}}{b} \; \left|\sum_m 2 C_m\left(A_m(\delta_s)+iB_m(\delta_s)\right) \right|^2
\end{eqnarray}
where
\begin{widetext}
\begin{eqnarray}
\label{Am}
A_m(\delta_s)=\left(2\text{arctan}\left(\frac{b/2+\nu_m}{\gamma_{m}}\right)+2\text{arctan}\left(\frac{b/2-\nu_m}{\gamma_{m}}\right)\right)-2\left(2\arctan\left(\frac{\left|\delta_s\right|+\nu_m}{\gamma_{m}}\right)+2\arctan\left(\frac{\left|\delta_s\right|-\nu_m}{\gamma_{m}}\right)\right)
\end{eqnarray}
\begin{eqnarray}
\label{Bm}
B_m(\delta_s)=\ln\left(\frac{(b/2+\nu_m)^2+\gamma_{m}}{(b/2-\nu_m)^2+\gamma_{m}^2}\right)-2\ln\left(\frac{(|\delta_s|+\nu_m)^2+\gamma_{m})^2}{(|\delta_s|-\nu_m)^2+\gamma_{m}^2}\right).
\end{eqnarray}
\end{widetext}
\begin{figure*} [ht]
   \begin{center}
\begin{picture}(500,230)
\put(0,0){\makebox(500,250){
\scalebox{0.55}[0.55]{
\includegraphics{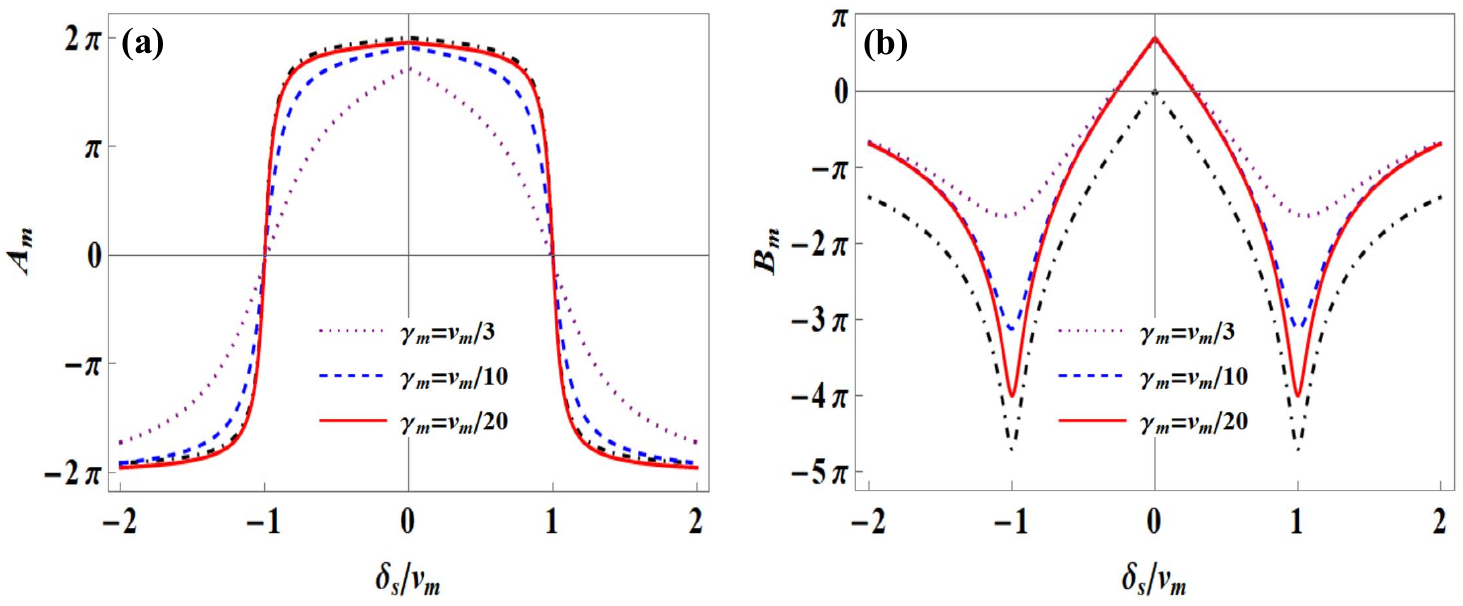}}}}
\end{picture}
\vspace{-2cm}
   \end{center}
\caption{\label{AB}  Dependence of the resonant contribution $A_m$ (a) and the off-resonant contribution $B_m$ (b) on the frequency $\delta_s$ of the phase flip for an intermediate level of frequency $\nu_m=b/4$, where $b$ is the bandwidth of the entangled input state. Sharp resonant features are obtained for a linewidth of $\gamma_m=\nu_m/20$ (solid line) and $\gamma_m=\nu_m/10$ (dashed line). At $\gamma_m=\nu_m/3$ (dotted line), the features are significantly reduced by the broadening of the resonance. For comparison, the dashed-dotted lines represent the corresponding dependence at a linewidth of $\gamma_m=\nu_m/20$ for the broadband limit. The main features of the dependence on the phase flip frequency $\delta_s$ do not change much when the bandwidth is increased.}
\end{figure*}
The dependence of the amplitudes $A_m(\delta_s)$ and $B_m(\delta_s)$ on the frequency $\delta_s$ of the phase flip is shown in Fig. \ref{AB}. The resonant contributions $A_m(\delta_s)$ describe a change of sign at $\delta_s=-\nu_m$ and at $\delta_s=+\nu_m$, broadened by the linewidth $\gamma_m$. The absolute value of this contribution only changes close to the resonance, where it drops to zero as the sign changes. This is very different from the dependence of the off-resonant contributions $B_m(\delta_s)$ on the frequency  $\delta_s$ of the phase flip. These contributions are rather close to zero at $\delta_s=0$ because of the cancellation of positive and negative contributions in the integral. The phase flip modifies this cancelation, achieving a maximal absolute value when the phase flip is at resonance ($\delta_s=\pm \nu_m$).

The precise resonant features described by Eqs.(\ref{Am}) and (\ref{Bm}) depend on the bandwidth $b$, the linewidth $\gamma_m$, and the resonant frequency $\nu_m$. In Fig. \ref{AB}, the variable $\delta_s$ has been scaled using the resonance $\nu_m$ to accommodate the fact that resonance is achieved at $\delta_s=\nu_m$. The linewidth then determines the broadening of the resonant features in a rather straightforward manner. The bandwidth $b$ has an effect on the values of $A_m$ and $B_m$ at $\delta_s=0$, which corresponds to a broadband entangled state without any phase flip.  It may be interesting to consider the values of $A_m$ and $B_m$ at $\delta_s=0$ in the limit of bandwidths that are much larger than $\nu_m$ and $\gamma_{m}$. In this broadband limit, $A_m(\delta_s=0)$ approaches $4 \arctan(\infty)=2\pi$ and $B_m(\delta_s=0)$ is negligibly small, reflecting the integrals of the real and imaginary parts of a Lorentz line over all frequencies. In general, the bandwidth $b$ introduces an offset to both $A_m$ and $B_m$, with positive values smaller than $2 \pi$ for $A_m$ and positive values that drop to zero in the broadband limit for $B_m$. In most of the following discussion, we will focus on the role of the linewidth $\gamma_m$ and the resonant frequency $\nu_m$, since the bandwidth dependent offset does not change the characteristic features of $A_m$ and $B_m$. Consistent with the observation that larger bandwidths are desirable when trying to achieve high off-resonant TPA rates \cite{Day04,Raymer2,SPIE21}, we find that the features introduced by the phase flips are easiest to observe in the broadband limit. However, sufficiently clear results can already be obtained when the bandwidth is given by $b=4 \nu_m$, which is the reason why we are using this bandwidth in Fig. \ref{AB}. For comparison, the broadband limit is shown for a linewidth of $\gamma_m=\nu_m/20$. As can be seen in the figure, the main features of the phase flip effects do not change much between a bandwidth of $b=4 \nu_m$ and the broadband limit.

At $\delta_s=0$, there is a significant resonant contribution $A_m$ and a much smaller contribution $B_m$. In the broadband limit, $A_m(\delta_s=0)=2 \pi$ and $B_m(\delta_s=0)=0$. As the phase flip frequency $\delta_s$ is moved towards resonance, $A_m$ decreases and $B_m$ increases, until the values at resonance are effectively reversed. For sufficiently narrow linewidths ($\gamma_m \ll \nu_m$), the broadband limit gives $A_m(|\delta_s|=\nu_m)=0$ and $B_m(|\delta_s|=\nu_m)=-4 \ln(2 \nu_m/\gamma_m)$, indicating an enhancement of the TPA rate given by the squared ratio of $B_m(|\delta_s|=\nu_m)$ and $A_m(\delta_s=0)$,
\begin{equation}
\label{eq:ratio}
\left(\frac{B_m(|\delta_s|=\nu_m)}{A_m(\delta_s=0)}\right)^2 \approx \left(\frac{2}{\pi}\ln(2 \nu_m/\gamma_m)\right)^2.
\end{equation}
However, this enhancement factor cannot be observed in isolation, since it always appears in a sum with all of the other intermediate levels $m$. For a systematic analysis of the enhancement effect, it is therefore convenient to consider the case of a single intermediate level first.
\section{Enhancement of absorption for a single intermediate resonance}
\label{sec:single}
\begin{figure*} [ht]
   \begin{center}
\begin{picture}(500,210)
\put(0,0){\makebox(495,240){
\scalebox{0.57}[0.57]{
\includegraphics{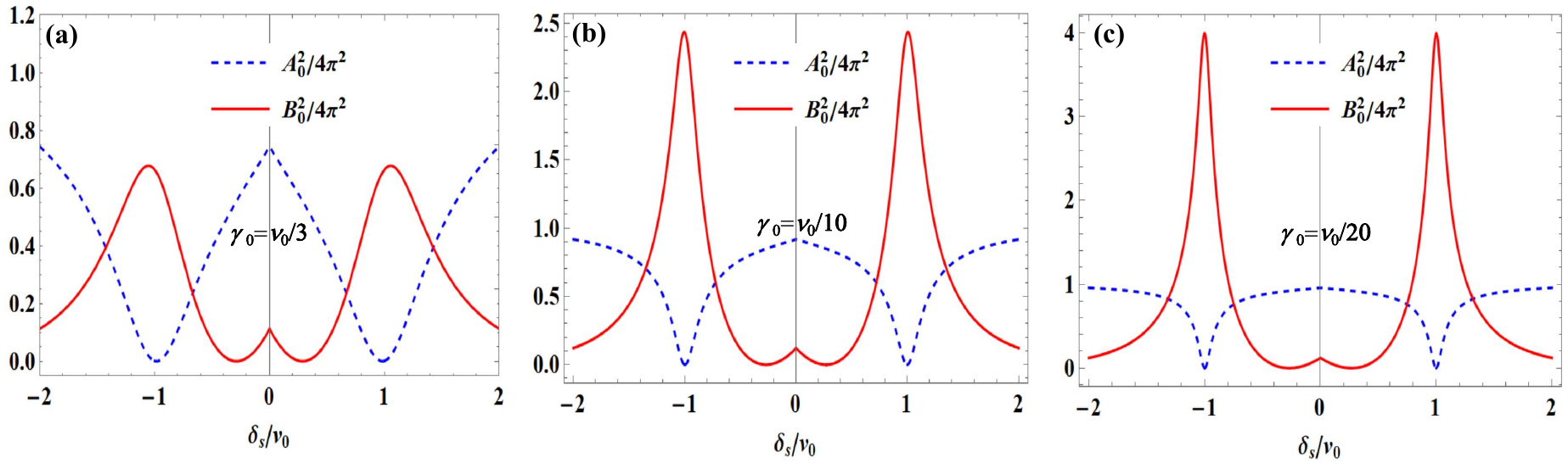}}}}
\end{picture}
\vspace{-2cm}
   \end{center}
\caption{\label{squares}  Contributions to the TPA rate for a single intermediate level for various linewidths $\gamma_0$ at a bandwidth of $b=4 \nu_0$. The dashed lines show the resonant contribution $A_0^2/4\pi^2$ and the solid lines show the off-resonant contribution $B_0^2/4\pi^2$ as a function of the frequency $\delta_s$ of the phase flip. (a) shows the results for a linewidth of $\gamma_0=\nu_0/3$, (b) shows the results for $\gamma_0=\nu_0/10$, and (c) shows the results for $\gamma_0=\nu_0/20$.}
\end{figure*}
In the case of a single intermediate level at a resonant frequency of $\nu_0$, the rate of TPA can be found by adding the squared values of the real and imaginary contributions $A_0$ and $B_0$,
\begin{eqnarray}
\label{single TPA}
P_{\mathrm{TPA}}(\delta_s)= \frac{P_{\mathrm{gf}}|C_0|^2}{b}(A_0^2(\delta_s)+B_0^2(\delta_s)).
\end{eqnarray}
To understand the physics of TPA enhancement, it is useful to consider the values of $A_0^2$ and of $B_0^2$ separately. Fig. \ref{squares} shows the resonant contribution $A_0^2$ and the off-resonant contribution $B_0^2$ for different linewidths. Using the squared values, it is easy to confirm the opposite roles of these two contributions. $B_0^2$ is very low at $\delta_s=0$ and initially drops to zero at small values of $|\delta_s|$, while $A_0^2$ is initially at its maximal value. As $\delta_s$ approaches the resonance at $|\delta_s|=\nu_0$, $A_0^2$ drops to zero and $B_0^2$ rises towards its maximum. It is important to note that the shape of the minimum of $A_0^2$ is not very different from the shape of the maximum of $B_0^2$, despite the different mathematical forms given by Eqs. (\ref{Am}) and (\ref{Bm}). As discussed above, the two contributions exchange roles on resonance, and enhancements are possible because the value of $B_m^2$ at resonance ($\delta_s=\nu_0$) exceeds the value of $A_m^2$ at $\delta_s=0$. However, Fig. \ref{squares} also illustrates the limits of this enhancement effect. At a linewidth of $\gamma_0=\nu_0/3$ and a bandwidth of $b=4 \nu_0$, the suppression of $A_0^2$ and the enhancement of $B_0^2$ approximately cancel each other. Enhancement thus depends on both bandwidth and linewidth, both expressed in terms of the separation $\nu_0$ between the resonance of the single intermediate level and the average photon frequency $\omega_{gf}/2$.

To evaluate the enhancement, it is useful to compare the TPA rate with a phase flip at $\delta_s$ with the TPA rate in the absence of a phase flip, corresponding to $\delta_s=0$,
\begin{equation}
\label{g}
g(\delta_s) = \frac{P_{\mathrm{TPA}}(\delta_s)}{P_{\mathrm{TPA}}(0)}.
\end{equation}

\begin{figure*} [ht]
   \begin{center}
\begin{picture}(500,360)
\put(0,0){\makebox(500,400){
\scalebox{0.6}[0.6]{
\includegraphics{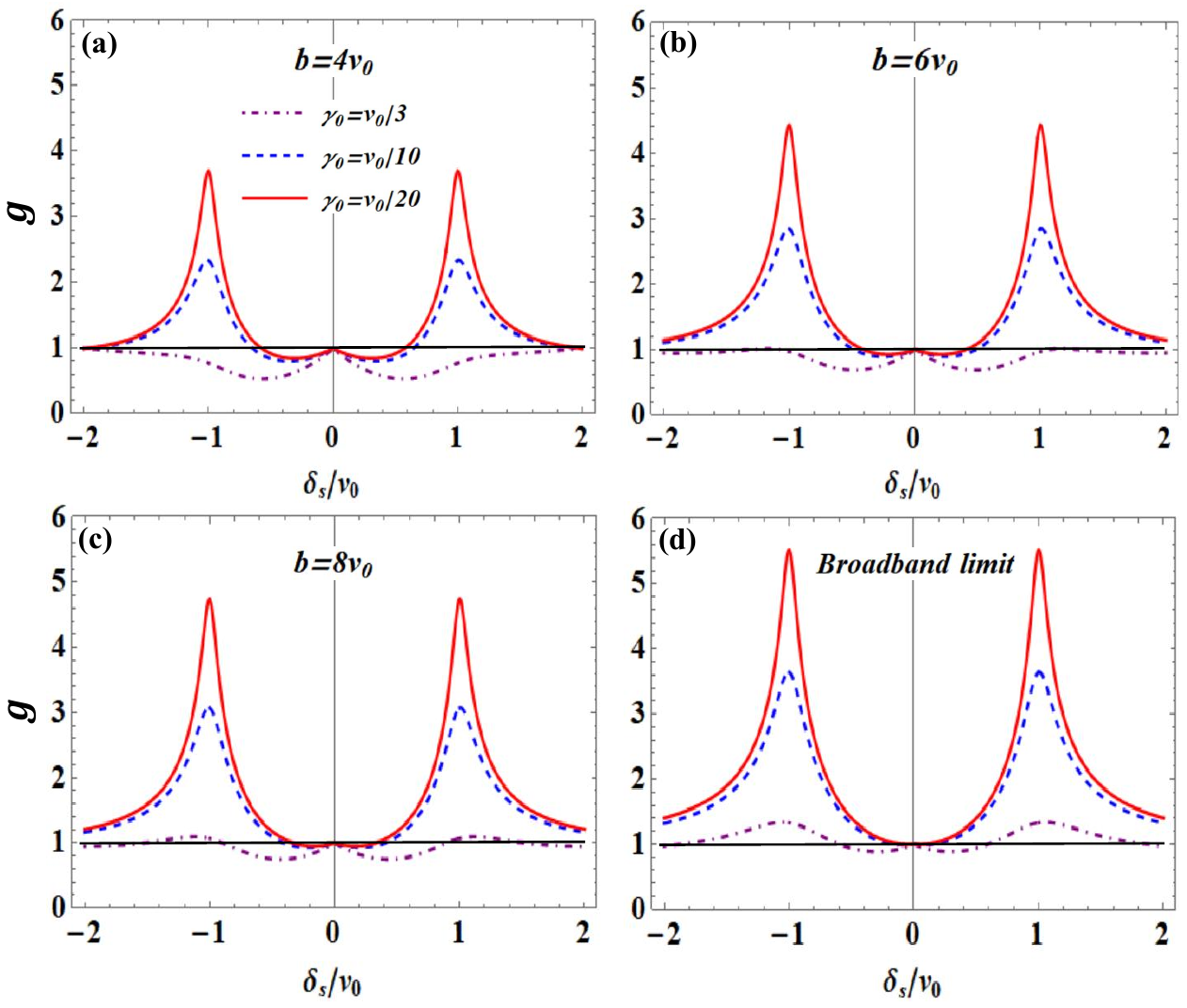}}}}
\end{picture}
\vspace{-2cm}
   \end{center}
\caption{\label{enhancement} The enhancement factor $g$ as a function of the frequency $\delta_s$ of the phase shift for different linewidths. (a) shows the enhancement for a bandwidth of $b=4\nu_0$, (b) shows the enhancement for $b=6\nu_0$, (c) shows the enhancement for $b=8\nu_0$, and (d) shows the enhancement for the broadband limit. The solid lines  shows the enhancement for $\gamma_0=\nu_0/20$, the dashed lines show the enhancement for $\gamma_0=\nu_0/10$, and the dashed-dotted lines show the enhancement for $\gamma_0=\nu_0/3$. The bold black line indicates $g=1$ (no enhancement) for reference.}
\end{figure*}

The enhancement factors for different linewidths are shown in Fig. \ref{enhancement}. The four panels show the enhancement factors for different bandwidths. In all cases, very little enhancement is observed at a linewidth of $\gamma_0=\nu_0/3$. At a bandwidth of $b=4\nu_0$, no enhancement is observed for the whole range of frequencies $\delta_s$. Even in the broadband limit, the TPA rate for $\gamma_0=\nu_0/3$ is actually reduced by phase flips close to $\delta_s=0$. The reason for this drop of the enhancement factor below one at small values of $|\delta_s|$ can be traced back to the reduction of the values of both $A_0^2$ and $B_0^2$ shown in Fig. \ref{squares}. However, significant enhancements are observed at narrower linewidths when the frequency of the phase flip is close to the resonance at $\delta_s=\pm \nu_0$. As expected from the previous discussion of $A_m$ and $B_m$, the enhancement effect becomes more pronounced as the bandwidth $b$ increases. Enhancements are observed for phase flips around $\delta_s=\nu_0$, and the maximal value of the enhancement depends on the ratio of linewidth $\gamma_0$ and resonant frequency $\nu_0$. Optimal results are obtained in the broadband limit, but the comparison between different bandwidths shows that the linewidth is the more relevant factor. Since the maximal enhancement is typically obtained at the resonant frequency $\delta_s=\nu_0$, it is useful to define the resonant enhancement as $g_{\mathrm{res.}}=g(\delta_s=\nu_0)$. In the broadband limit, the corresponding resonant enhancement factor is given by
\begin{widetext}
\begin{equation}
\label{eq:maxg}
g_{res.}(\mathrm{broadband})=(1 - \frac{2}{\pi}\arctan(2\nu_0/\gamma_0))^2+(\frac{2}{\pi} \ln(1+2 \nu_0/\gamma_0))^2.
\end{equation}
\end{widetext}

\begin{figure} [ht]
   \begin{center}
\begin{picture}(500,250)
\put(0,0){\makebox(500,270){
\scalebox{0.5}[0.5]{
\includegraphics{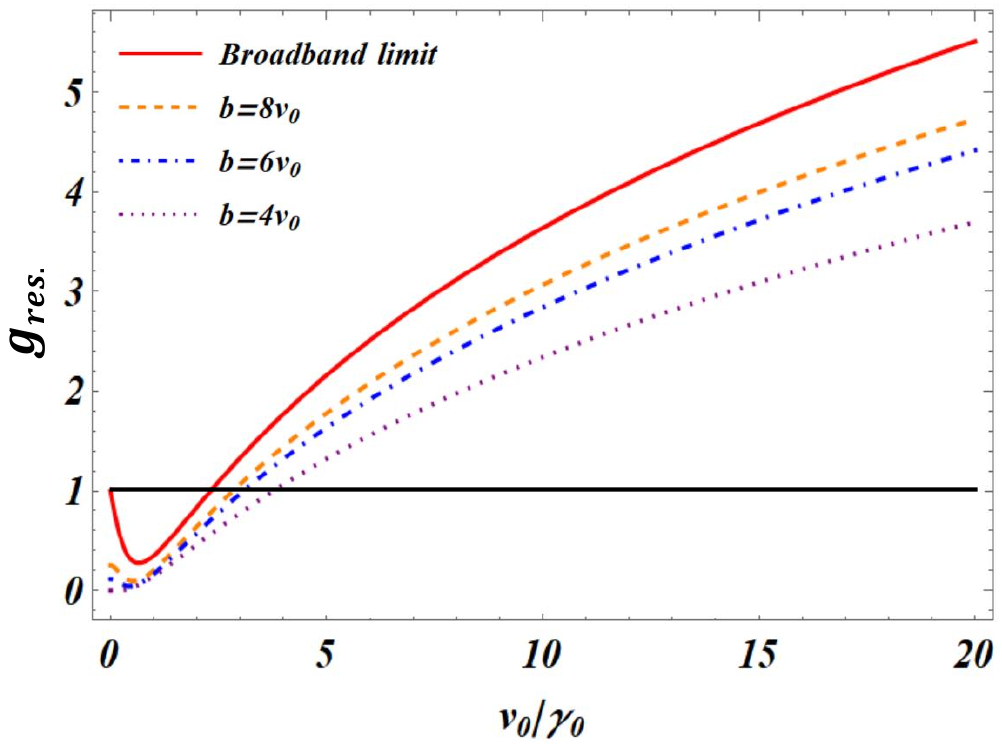}}}}
\end{picture}
\vspace{-2cm}
   \end{center}
\caption{\label{compare} Dependence of the resonant enhancement factor $g_{res.}$ on the inverse linewidth $\nu_0/\gamma_0$. The solid line gives the broadband limit, the dashed line corresponds to a bandwidth of $b=8 \nu_0$, the dashed-dotted line to a bandwidth of $b=6 \nu_0$, and the dotted line to a bandwidth of $b=4 \nu_0$. The bold black line indicates $g_{res.}=1$ (no enhancement) for reference.}
\end{figure}

Fig. \ref{compare} shows the dependence of resonant enhancement on the ratio of resonant frequency $\nu_0$ and linewidth $\gamma_0$. Since we are mostly interested in the conditions under which large enhancements can be achieved, we focus on the regime of narrow linewidths above $\nu_0/\gamma_0=5$, where the enhancement factors are larger than one. The broadband limit describes the maximal enhancement for the respective linewidth ratio, and the enhancements achieved at narrower bandwidths are correspondingly lower. As discussed previously, this is a direct result of the non-zero value of the off-resonant contribution $A_0^2$ at $\delta_s=0$. The lower the bandwidth, the narrower the linewidth must be to achieve a given enhancement. However, the qualitative dependence of resonant enhancement on the linewidth ratio is independent of the bandwidth. Even in the broadband limit, enhancement is only observed when the linewidth is sufficiently narrow.

For sufficiently narrow linewidths, it is possible to simplify the broadband limit formula by using Eq.(\ref{eq:ratio}). The approximate formula is given by
\begin{equation}
\label{rough g}
g_{res.}(\mathrm{broadband})\approx (\frac{2}{\pi} \ln(2 \nu_0/\gamma_0))^2.
\end{equation}
This formula is particularly useful to estimate the necessary inverse linewidth for a specific enhancement factor. For instance, an enhancement by a factor of $4$ requires a minimal inverse linewidth of $\nu_0/\gamma_0 = \frac{1}{2} e^{\pi}$, a ratio of approximately $11.6$. Although the formula is less reliable at lower enhancement factors, it may be interesting to use it to estimate the inverse linewidth required for an enhancement factor of one. The formal result is $\nu_0/\gamma_0 =\frac{1}{2} e^{\pi/2}$, a ratio of approximately $2.4$. Although by no means precise, it seems reasonable to assume that no enhancement can be observed when the inverse linewidth $\nu_0/\gamma_0$ is smaller than $2.4$.

At first sight, it may seem difficult to find materials with sufficiently narrow linewidths for a significant TPA enhancement using a resonant phase flip. However, it should be kept in mind that the condition refers to the ratio of the linewidth $\gamma_0$ and the resonant frequency $\nu_0$. If $\nu_0$ is large, it may be possible to satisfy the condition even if $\gamma_0$ is relatively broad. Ideally, we would be looking for materials with intermediate levels in the center of the lower or upper half of the broadband spectrum, where $\nu_0=b/4$. If the bandwidth of the input state is sufficient, the resonant band at $\nu_0$ could be rather wide and still produce a resonant enhancement effect. If a scan of the phase flip frequency is used as a method of spectroscopy, it may therefore be possible to use this enhancement effect to identify intermediate levels with resonant transitions at very high or very low frequencies. The argument that was given here for a single resonant level may then be applied to a whole band of intermediate states that are too close to each other to be resolved. It may be good to remember that the introduction of broadened resonances was motivated by the dephasing and relaxation effects in open systems. We can thus expect that a number of complicated systems can be described in this approximate manner. To understand the possible limitations of this method, it is necessary to consider the possible effects of interferences between different intermediate levels, as described by the general formula in Eq. (\ref{P_AB}).

\section{Interference effects between different intermediate resonances}
\label{sec:Interference}

The basic enhancement effect consisting in an increase of the value of $B_m$ at resonance can be observed even in the presence of multiple intermediate levels $m$. However, it needs to be remembered that the sign of $A_m$ changes at resonance, resulting in a switch between constructive and destructive interferences of $A_m$ with the contributions from other intermediate levels. In this section, we will investigate the effects of such interferences on the TPA rate.

As an example, let us consider the case of two intermediate levels at resonances $\nu_1$ and $\nu_2$ with coefficients of $C_1$ and $C_2$ describing the associated transition matrix elements. According to Eq.(\ref{P_AB}), the contributions of the two levels must be added before the absolute square determines the TPA rate,
\begin{eqnarray}
\label{interference}
P_{\mathrm{TPA}}(\delta_s)=\frac{4P_{\mathrm{gf}}}{b} \; \left| C_1\left(A_1+iB_1\right)+C_2 \left(A_2+iB_2\right) \right|^2.
\end{eqnarray}
To analyze the interference effects, it is necessary to consider the phase relation between the coefficients $C_1$ and $C_2$. For simplicity, we will only consider real values of $C_1$ and $C_2$, with positive signs describing constructive interferences and opposite signs describing destructive interferences. It may be worth noting that other phase differences would introduce a rather complicated structure of interferences involving all possible combinations of $A_1$ and $B_1$ with $A_2$ and $B_2$. The advantage of limiting ourselves to real coefficients is that we can separate the interference between $A_1$ and $A_2$ from the interference between $B_1$ and $B_2$. If the absolute values of $C_1$ and $C_2$ are the same, interference effects are described by the separate sums and differences of $A_1$ and $A_2$, and of $B_1$ and $B_2$.

\begin{figure*} [ht]
   \begin{center}
\begin{picture}(500,340)
\put(0,0){\makebox(500,370){
\scalebox{0.6}[0.6]{
\includegraphics{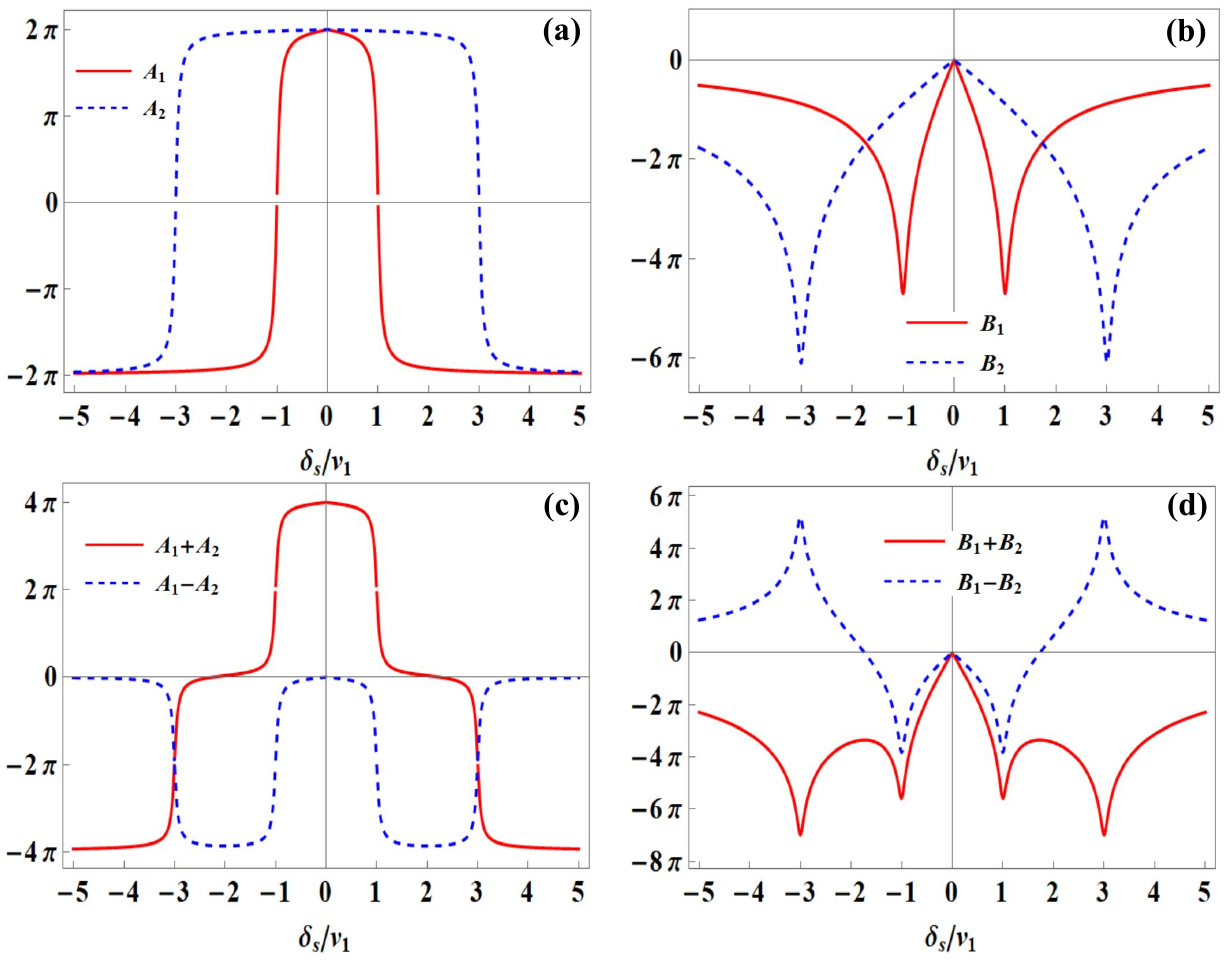}}}}
\end{picture}
\vspace{-2cm}
   \end{center}
\caption{\label{onetwo} Illustration of the effects of interferences between two levels for $\nu_2=3 \nu_1$ at a linewidth of $\gamma_1=\gamma_2=\nu_1/20$. (a) shows the $\delta_s$-dependence of resonant contributions $A_1$ (solid line) and $A_2$ (dashed line), (b) shows the $\delta_s$-dependence of off-resonant contributions $B_1$ (solid line) and $B_2$ (dashed line), (c) shows the constructive and destructive interferences $A_1+A_2$ (solid line) and $A_1-A_2$ (dashed line), and (d) shows the constructive and destructive interferences $B_1+B_2$ (solid line) and $B_1-B_2$ (dashed line).}
\end{figure*}

Fig.\ref{onetwo} illustrates the basic interference effects between two intermediate levels with the same linewidths $\gamma_1=\gamma_2=\nu_1/20$, where the resonances are sufficiently separate to recognize two distinct enhancement effects. Interferences between $A_1$ and $A_2$ result in regions where the two contributions cancel and regions where they add up. The contributions of $A_1+A_2$ cancel each other for $\nu_1<|\delta_s|<\nu_2$ and add up outside of this region. For $A_1-A_2$, the contributions add up for $\nu_1<|\delta_s|<\nu_2$ and cancel each other outside of this region. $B_1$ and $B_2$ are dominated by the resonant enhancements. The effects of interferences are strongest for $\nu_1<|\delta_s|<\nu_2$, where $B_1-B_2$ drops to zero and changes its sign, whereas $B_1+B_2$ remains at high absolute values throughout. In summary, constructive interference results in negligible values of $A_1+A_2$ and high values of $B_1+B_2$ in the region between the resonances, while destructive interferences result in high values of $A_1-A_2$ and low values of $B_1-B_2$. In the sum of the squares, these two interference effects may be hard to distinguish.

\begin{figure*} [ht]
   \begin{center}
\begin{picture}(500,220)
\put(0,0){\makebox(500,250){
\scalebox{0.62}[0.62]{
\includegraphics{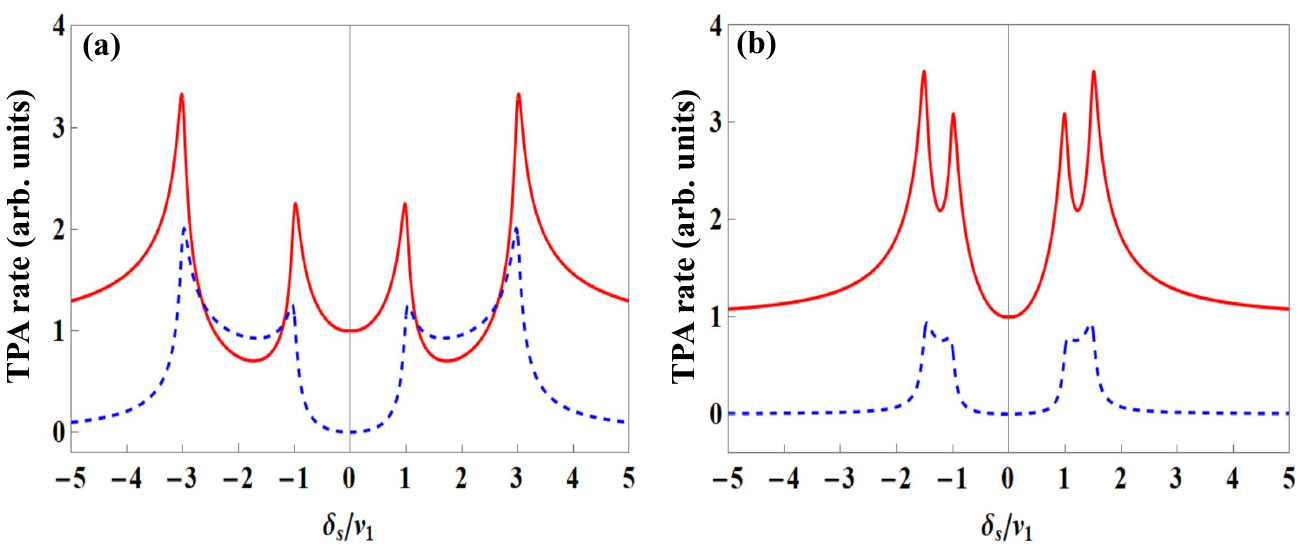}}}}
\end{picture}
\vspace{-2cm}
   \end{center}
\caption{\label{fig6} Dependence of TPA rates of interferences between two intermediate levels on the frequency $\delta_s$ of the phase flip for the same linewidths of $\gamma_1=\gamma_2=\nu_1/20$. The  solid lines show the results for constructive interference and the dashed  lines show the results for destructive interference. In (a) the second resonance is at $\nu_2=3 \nu_1$ and in (b) it is at $\nu_2=1.5 \nu_1$.}
\end{figure*}

The characteristic difference between constructive and destructive interference is seen at $\delta_s=0$, corresponding to the TPA rate without a phase flip. In the broadband limit, we have $A_1(0)=A_2(0)=2\pi$ and $B_1(0)=B_2(0)=0$. Constructive interference results in $A_1(0)+A_2(0)=4 \pi$ and destructive interference results in $A_1(0)-A_2(0)=0$. Due to this interference effect, it is not useful to compare the TPA rates at different frequencies $\delta_s$ with the TPA rate at $\delta_s=0$. Instead, we will simply compare the two results directly. Fig. \ref{fig6} shows the TPA rates for constructive and destructive interferences, with all other parameters being equal. As expected, destructive interference reduces the resonant TPA rates while constructive interferences enhance them. However, destructive interference completely suppresses the TPA rates close to $\delta_s=0$. It is therefore quite remarkable that comparatively high TPA rates can be obtained for destructive interferences when the frequency of the phase flip is between the resonances at $\nu_1<|\delta_s|<\nu_2$. This activation of transitions that interfere destructively may be one of the most interesting aspects of the phase relation between the coefficients $C_m$. Constructive interference is consistent with the idea that levels close to each other correspond to a single broadened transition, even though the individual resonances can be resolved as long as the linewidths are sufficiently narrow. The case of destructive interference requires a more thorough analysis due to the suppression of TPA rates at $\delta_s=0$.


\section{Conclusions}
\label{sec:conclude}

We have shown that a spectral phase flip introduced at a frequency difference of $\delta_s$ from the average frequency of the entangled photons can enhance the TPA of the entangled photons significantly when it is resonant with an intermediate level involved in the TPA process. This effect might be useful in the characterization of intermediate levels far away from the average photon frequency, providing access to frequencies at the far ends of the broadband spectrum. Interference effects between different transitions might complicate the picture somewhat, with an interesting possibility of activating forbidden transitions when the phase flip is placed between two destructively interfering transitions. In general, the application of a spectral phase flip requires no prior knowledge of the level structure of a two-photon absorber and can be adjusted for optimal enhancement effects based on the experimental data. The dependence of TPA rates on the frequency of the phase flip highlights a fundamental aspect of phase dispersion in the TPA process that may have a wide range of applications in the future.

\begin{acknowledgments}
Baihong Li was supported by National Natural Science Foundation of China (12074309) and the Youth Innovation Team of Shaanxi Universities. Holger F. Hofmann was supported by JST-CREST (JPMJCR1674), Japan Science and Technology Agency.
\end{acknowledgments}

\end{CJK*}
\end{document}